\documentclass[%
 reprint,
 amsmath,amssymb,
 aps,
prb,
]{revtex4-2}

\usepackage{graphicx}
\usepackage{dcolumn}
\usepackage{bm}

\begin{document}


\title{Peculiarities of spin- orbit interaction systematically measured in  FeCoB nanomagnets}

\author{Vadym Zayets}
\affiliation{National Institute of Advanced Industrial Science and Technology (AIST), Umezono 1-1-1, Tsukuba, Ibaraki, Japan}

\date{\today}

\begin{abstract}
This study introduces a method to measure strength of spin-orbit interaction (SO) in a nanomagnet,  investigating fundamental phenomena governing magnetic anisotropy. The method explores the fundamental property of SO  in its linear proportionality to the external magnetic field, a relationship validated through experimental observation. Systematic study of SO in FeCoB nanomagnets reveals distinct SO behaviors in bulk and at an interface, its substantial disparities in single- and multi-layer nanomagnets, intriguing periodic oscillations in SO strength, and the systematic relationship between SO strength, demagnetization field, and magnetic anisotropy based on surface imperfections.  These findings provide crucial insights into diverse spin-orbit interaction behaviors, crucial for understanding and optimizing magnetic anisotropy and nanomagnet properties.
\end{abstract}


\maketitle

Spin-orbit interaction (SO) refers to a magnetic field $H_{SO}$ of relativistic origin \cite{LandauField,Dirac}  experienced by an electron while moving within an electric field $E$:

\begin{equation}
	H_{so}= \frac{v}{c^2}  E
	\label{SO_basic}
\end{equation}

where $v$ is a component of the  electron velocity perpendicular to $E$.

 Within a nanomagnet, the spin-orbit interaction plays a crucial role in establishing the magnetic anisotropy, which effectively maintains the nanomagnet's magnetization along its easy magnetic axis \cite{Johnson1996Hani,MMM2020_Hani_Hoff}. This magnetic anisotropy leads to two stable magnetization directions within the nanomagnet, allowing data to be stored using these stable states. As a result, a nanomagnet becomes valuable for non-volatile memory applications \cite{MRAMCubukcu,MRAM2017Ohno,MRAM2020Everspin}.

As the density of magnetic memory increases, the size of nanomagnets decreases accordingly. Consequently, the magnetic energy diminishes and reaches a point where it becomes comparable to thermal energy \cite{MRAMCubukcu,MRAM2017Ohno,MRAM2020Everspin,ZayetsArch2019Hc}. In such a scenario, thermal fluctuations can lead to magnetization reversal, resulting in data loss within the magnetic memory. To prevent this undesirable occurrence, it is essential to enhance the strength of the magnetic anisotropy in nanomagnets, which can be achieved through the optimization and reinforcement of the spin-orbit interaction.

In amorphous ferromagnetic metals such as FeB and FeCoB, which are mainly used as a material for MRAM, the magnetic anisotropy originates primarily at the interface \cite{MRAMCubukcu,MRAM2017Ohno,MRAM2020Everspin}. The orbits of localized electrons at the interface undergo significant deformation towards the neighboring atomic layer. This substantial deformation leads to a remarkable enhancement in the strength of the spin-orbit interaction and results in perpendicular magnetic anisotropy (PMA). Due to this remarkable interface enhancement of magnetic anisotropy,  the magnetic energy of a nanomagnet significantly increases, making the nanomagnet  thermally stable even at smaller sizes. Consequently, a nanomagnet with PMA becomes the preferred choice for magnetic memory applications \cite{MRAMCubukcu,MRAM2017Ohno,MRAM2020Everspin}.

As seen in Eq. \ref{SO_basic}, $H_{so}$ is proportional to $1/c^2$ and $E$. Therefore, the magnetic field  of the spin-orbit interaction $H_{so}$ is only significant when the electron moves through a large electric field at a speed close to a substantial fraction of the speed of light $c$. This occurs during the orbital movement around the atomic nucleus. However, certain conditions must be met for the spin-orbit interaction to be large in the case of orbital movement.  The key requirement is the breaking of time-inversion symmetry (T-sym), as without such breaking, there would be no spin-orbit interaction (SO). For instance, the orbital movement of the electron can be separated into clockwise (CW) and counterclockwise (CCW) components, each contributing to $H_{so}$. In the absence of T-sym breaking, $H_{so}$ generated by the CW and CCW components are of equal strength but in opposite directions, resulting in a cancellation, leading to no overall $H_{so}$. 

The time-inversion symmetry is broken for an electron having a non-zero orbital moment. For instance, conduction electrons possess an orbital moment, leading to a substantial contribution to its energy  from the spin-orbit interaction \cite{Luttinger1955}. Another method of breaking T-sym is by applying an external magnetic field  $H_{ext}$. This magnetic field induces a Lorentz force acting in opposite directions for the CW and CCW components of the wavefunction. Consequently, the Lorentz force leads to a modification of the wavefunction, causing one component to, on average, move closer to the nucleus while the other component moves away from it. This imbalance in the contributions from these components to the Spin-Orbit Interaction results in the creation of magnetic field $H_{so}$. Since in the absence of $H_{ext}$, T-sym remains unbroken and $H_{so}$ equals zero, a linear relationship between $H_{so}$ and $H_{ext}$ can be assumed:

\begin{equation}
	H_{so}=k_{so} \cdot H_{ext}
	\label{HsoVsH}
\end{equation}

where $k_{so}$ is the coefficient of spin- orbit interaction.

In situations where the electron's orbital lacks symmetry, the strength of the spin-orbit interaction (SO) becomes dependent on the orientation of the external magnetic field $H_{ext}$. The SO magnetic field $H_{so}$ arises due to the imbalance created by the Lorentz force between two components of the orbital, and, consequently,  the initial shape and symmetry of the electron orbital substantially influence SO strength. This implies that the value of $k_{so}$ is orientation-dependent, particularly concerning the direction in which the magnetic field is applied relative to the electron orbital's asymmetry. 

The amplification of SO occurs when the electron orbital is in closer proximity to the atomic nucleus, where the nucleus's electrical field is most potent (Eq. \ref{SO_basic}). This proximity is achieved when the center of the orbital deviates from the nucleus's position. Such a deviation can occur in the vicinity of an interface where the electron interacts with two distinct types of atoms on either side, prompting the necessary shift. Consequently, the enhancement of SO for an electron at the interface predominantly occurs when the magnetic field is applied perpendicular to the interface. Even when the magnetic field is applied parallel to the interface, SO remains present but with a smaller $k_{so}$.

The creation of the magnetic anisotropy in a ferromagnetic material is the key feature of the spin- orbit interaction \cite{Johnson1996Hani,MMM2020_Hani_Hoff}. Inside a ferromagnet, there exists an intrinsic magnetic field aligned with the direction of ferromagnet magnetization. The intrinsic magnetic field creates $H_{so}$ as described by Eq.\ref{HsoVsH}.  Since $k_{so}$ is larger  in the direction perpendicular- to -plane ($\perp$-plane),  $H_{so}$ larger and magnetic energy higher in the $\perp$-plane direction compared to the in-plane direction. This implies that the spin-orbit interaction  makes the $\perp$-plane magnetization direction energetically favorable.

Another factor influencing the equilibrium magnetization direction in a ferromagnetic nanomagnet is demagnetization. The demagnetization field is oriented in the $\perp$-plane direction and opposes the direction of magnetization. The demagnetization field effectively lowers the magnetic energy in the $\perp$-plane direction. Consequently, in the absence of spin-orbit interaction, the equilibrium magnetization direction is the in-plane direction \cite{Johnson1996Hani}.

The presence of the Spin-Orbit Interaction counteracts the demagnetization effect and promotes the $\perp$-plane direction as the equilibrium magnetization orientation. For instance, this phenomenon is observed in thin amorphous nanomagnets, where the substantial spin-orbit interaction exists only for electrons at  the interface. In the case of a thicker nanomagnet, the bulk contribution dominates, the demagnetization effect overcomes SO interaction and the equilibrium magnetization direction is in-plane. Conversely, for a thinner nanomagnet, the interface contribution dominates, the SO interaction counteracts the demagnetization effect  and the equilibrium magnetization direction becomes $\perp$-plane \cite{Ohno2010PMA_Thickness}.

Since the strength of the magnetic anisotropy is inherently linked to the strength of the spin-orbit interaction, the most intuitive and direct approach for a measurement of the SO strength is through measurements of magnetic anisotropy. This approach further benefits from the linear dependence of SO strength on the external magnetic field (Eq.\ref{HsoVsH}). Consequently, by measuring the dependence of the strength of magnetic anisotropy on the external magnetic field, it becomes possible to discern and quantify the contribution of the spin-orbit interaction. In this study, we have developed and employed this measurement method to systematically evaluate and study the strength of SO in FeCoB nanomagnets.

  To measure the strength of magnetic anisotropy, an external magnetic field $H_x$ is applied perpendicular to the easy magnetic axis, causing the magnetization to tilt. The strength of the magnetic anisotropy is evaluated from the magnetization tilt. There is an intrinsic magnetic field in a nanomagnet, which holds magnetization along its easy axis. The intrinsic magnetic field is larger in a nanomagnet with a stronger magnetic anisotropy. Consequently, as the intrinsic magnetic field strengthens for a stronger magnetic anisotropy, a stronger perpendicular field $H_x$ becomes necessary to induce an equivalent tilt in the magnetization. The external magnetic field $H_x$, which  aligns the magnetization fully along  the hard magnetic axis, is referred to as the anisotropy field $H_{ani}$. $H_{ani}$ serves as a descriptor of the magnetic anisotropy strength. In this study, the tilt of the  nanomagnet magnetization was measured by a measurement of the Hall angle. For this purpose, the FeCoB nanomagnets were fabricated on top of a Ta- or W- made Hall bar (See Appendix \ref{AppendixExpSetup}).
  
  \begin{figure}[t]
  	\begin{center}
  		\includegraphics[width=6.5cm]{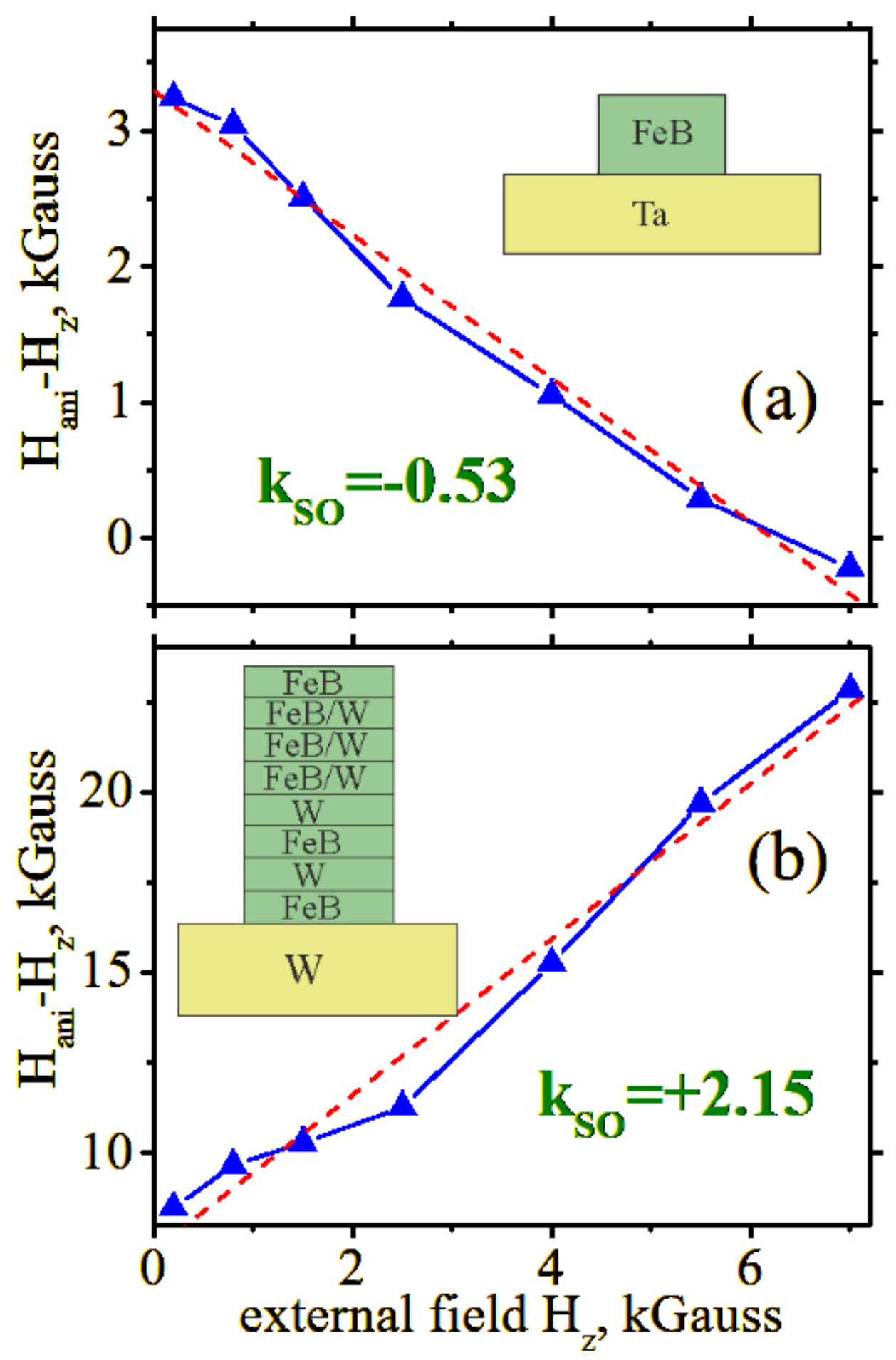}
  	\end{center}
  	\caption{\label{fig:fig1} 
  		Anisotropy field $H_{ani}$ vs. external magnetic field $H_{z}$ measured (a) in $Ta(2):FeB(1.1)$ and (b) $W(3):[FeB(0.55):W(0.5)]_{5}:FeB(0.55)$ nanomagnets. The slope of the dependence is proportional to SO strength and  gives the coefficient of SO interaction.  
  	}
  \end{figure}
  
  
   There exists a linear relationship between the in-plane component of magnetization $M_x$ and $H_x$ (See Appendix \ref{AppendixHaniCalculation}):

\begin{equation}
	\frac{M_x}{M}=\frac{H_x}{H_{ani}}
	\label{Hani}
\end{equation}

  By performing a linear fit of the measured $M_x/M$ against $H_x$, the value of $H_{ani}$ was evaluated. Additionally, an external magnetic field $H_z$ was applied perpendicular to the plane, aligning with the easy magnetic axis, serving as a parameter for the measurement. For each $H_z$ value, an in-plane field $H_x$ was systematically scanned, enabling the evaluation of $H_{ani}$ vs. $H_z$. Equilibrium magnetization direction of all measured nanomagnet is $\perp$-plane.

 The equilibrium magnetization direction  under an applied  tilted magnetic field was calculated in Appendix \ref{AppendixHaniCalculation} by minimizing the magnetic energy. Subsequently, it gives the dependence of the anisotropy field $H_{ani}$ on the external magnetic field $H_z$ as follows:

\begin{equation}
	H_{ani}= H^0_{ani}+H_z+k_{so}H_z
\end{equation} 

where $H^0_{ani}$ is the anisotropy field in absence of $H_z$.

As anticipated, $H_{ani}$ exhibits a linear proportionality to $H_z$. The second term in the equation corresponds to the bulk contribution, representing the alignment of magnetization along $H_z$, which doesn't provide informative insights. Consequently, when conducting data analysis, it is more advantageous to utilize the relationship between  $H_{ani}-H_z$, and  $H_z$. This approach ensures that the significant bulk contribution does not obscure a potentially weaker dependence on $k_{so}$.

\begin{figure}[t]
	\begin{center}
		\includegraphics[width=6.5cm]{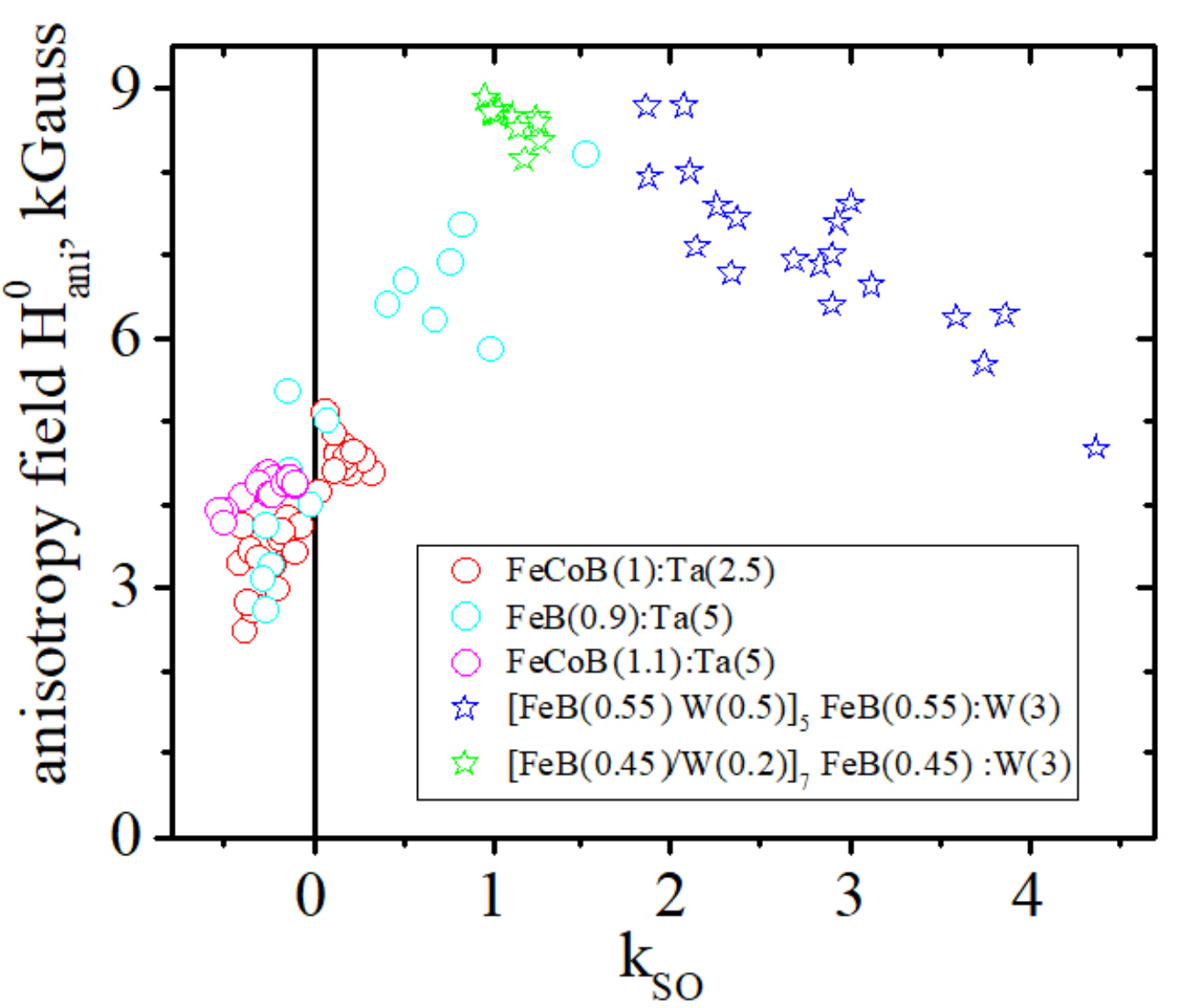}
	\end{center}
	\caption{\label{fig:fig2} 
		Measured anisotropy field $H^0_{ani}$ vs. the measured coefficient of spin-orbit interaction $k_{so}$ in nanomagnets of varying sizes and structures. Each dot represents an individual nanomagnet measurement. Dots of the same color and shape correspond to nanomagnets fabricated at different locations on the same wafer. Stars indicate multilayer nanomagnets, while circles represent single-layer nanomagnets. 
	}
\end{figure}

Figure \ref{fig:fig1} displays the measured relationship between $H_{ani}-H_z$ and the external magnetic field $H_z$ in a single-layer nanomagnet and a multi-layer nanomagnet. In both cases, the relationship exhibits an approximately linear trend, but the slopes differ significantly. For the single-layer nanomagnet, the slope is negative, while for the multi-layer nanomagnet, it is positive. It means that the strength of the spin-orbit interaction is opposite for those nanomagnets. $k_{so}$ is negative for the single-layer nanomagnet but positive for the multi-layer nanomagnet. This divergence is a consistent pattern. In Figure \ref{fig:fig2}, the measured values of $H^0_{ani}$ are plotted against the measured values of $k_{so}$ for nanomagnets of various sizes and structures. In all multi-layer nanomagnets, $k_{so}$ consistently exhibits a large and positive value. In contrast, for single-layer nanomagnets, $k_{so}$ is generally smaller and can be either positive or negative. A negative value of $k_{so}$ indicates that, when considering an average over bulk and interface contributions, the strength of the spin-orbit interaction in the in-plane direction is greater than in the $\perp$-plane direction (See Eq. \ref{kso_effective}).

The significant disparity in strength of spin-orbit interaction between single- and multi-layer nanomagnets underscores the substantial influence of interfaces on the overall strength. A greater number of interfaces correlates with a larger $k_{so}$. Moreover, each interface incrementally adds to the overall strength, irrespective of the interface's polarity. For instance, both Pt/FeB and FeB/Pt interfaces enlarge the overall strength. That is the reason why  $k_{so}$ in a multi- layer nanomagnet  is substantially larger than in a single- layer nanomagnet.

Comparative analysis of the measured $k_{so}$ in single- and multi-layer nanomagnets reveals that the interface contribution to $k_{so}$ is positive and large. Conversely, the contribution from the bulk of the nanomagnet is small  and negative. In single-layer nanomagnets, there's a balancing between the bulk and interface contributions, resulting in a small $k_{so}$ that can be either positive or negative.

It indicates a notable distinction in the deformation of Fe and Co orbitals between the interface and the bulk regions. In the bulk, orbital deformation remains minimal, resulting in an almost isotropic tensor $\hat{k}_{so}$. Elements $k_{so,x}$ and $k_{so,y}$ are slightly larger than $k_{so,z}$, which contributes to an overall small and negative effective $k_{so}$ within the bulk.

Conversely, at the interface, Fe and Co orbitals undergo significant deformation along the interface's normal. This deformation arises from differing orbital interactions with orbitals situated below and above the interface. Consequently, the tensor $\hat{k}_{so}$ assumes a uniaxial orientation, where $k_{so,z}$ becomes notably larger than the in-plane elements $k_{so,x}$ and $k_{so,y}$. This configuration gives rise to a substantial and positive effective $k_{so}$ for the orbitals positioned at the interface (See Eq. \ref{kso_effective}).

Additional observation in Fig. \ref{fig:fig2} shows scattering among the data points of nanomagnets fabricated at different locations on the same wafer. Some wafers display data convergence within a small area, while others exhibit data spreading over a larger range. A more homogeneous and smoother wafer corresponds to reduced data spreading. This spread in data serves as an indicator of fabrication technology perfection. Both variations in thickness and surface roughness contribute to data dispersion on a wafer. Given that nanomagnet thickness is around 1 nm, even a variance within a single atomic layer range markedly impacts $k_{so}$ and $H^0_{ani}$. 

Moreover, the surface's perfection plays an even more substantial role, exerting a notable influence on $k_{so}$ and $H^0_{ani}$. As previously discussed, robust spin-orbit interaction at an interface arises from substantial disparities in electron interactions with neighboring orbitals positioned above and below the interface. This leads to substantial and asymmetric orbital deformation along the surface normal. Deterioration in interface smoothness and sharpness weakens this deformation, causing its direction to randomly incline  from the interface normal, resulting in a reduced $k_{so}$. Consequently, the value of $k_{so}$ acts as an indicator of interface smoothness and sharpness, providing another insight into the quality of fabrication technology. Keeping other nanomagnet properties constant, a higher $k_{so}$ value signifies a smoother and sharper interface.

The data from nanomagnets from a single wafer in Fig. \ref{fig:fig2} don't spread randomly; instead, they follow a linear trend. In the case of single-layer nanomagnets, the slope of this trend line is consistently positive. However, for multi-layer nanomagnets, the slope is primarily negative, turning positive only for nanomagnets with a minimal number of layers. The correlation between variations in $H^0_{ani}$ and $k_{so}$ is expected because $H^0_{ani}$ is linearly proportional to $k_{so}$ (see Eq.\ref{HaniVsHint}). Thus, the variation in data for $H^0_{ani}$ vs. $k_{so}$ should follow a line, and the slope of this line should be positive. The existence of a negative slope indicates that, in addition to $k_{so}$, there is another parameter affecting $H^0_{ani}$, and its influence is opposite to that of $k_{so}$. This is why the slope polarity may change from positive to negative.

The parameter that satisfies these properties is demagnetization. Demagnetization is the magnetic field created by the magnetic dipole formed at opposite surfaces of a magnetic film. The direction of the demagnetization field $H_{demag}$ is opposite to the direction of the internal magnetic field, thus reducing $H^0_{ani}$ (see Eq. \ref{HaniVsHint}). Surface roughness and diminished interface sharpness reduce the demagnetization field. Interface roughness introduces random deviations in the demagnetization direction from the interface normal at each local point, resulting in the effective reduction of the average demagnetization field. A variation in interface roughness and sharpness leads to variations in the $H_{demag}$  for nanomagnets fabricated at different locations on the same wafer. This contributes to the data spread observed in Fig. \ref{fig:fig2}.

In total, a smoother interface makes $k_{so}$ larger, leading to a larger $H^0_{ani}$. However, a smoother interface also enhances $H_{demag}$, which conversely reduces $H^0_{ani}$. Therefore, there exists a delicate balance between these two effects. For a single-layer nanomagnet, the first effect prevails, resulting in the positive slope seen in Fig.\ref{fig:fig2}. Conversely, for multi-layer nanomagnets, the impact of the demagnetization field prevails, yielding the negative slope. The reason for the negative slope and, therefore, the dominance of $H_{demag}$  can be explained as follows: Both $k_{so}$ and $H_{demag}$ are large in a multi- layer nanomagnet. As the demagnetization field becomes nearly equal to the magnetization's field, the intrinsic magnetic field within the nanomagnet becomes small.  Consequently, even a slight variation in interface roughness and subsequently the demagnetization field induces a notable relative change in the already small intrinsic magnetic field, having a large effect on the anisotropy field.

\begin{figure}[t]
	\begin{center}
		\includegraphics[width=6.5cm]{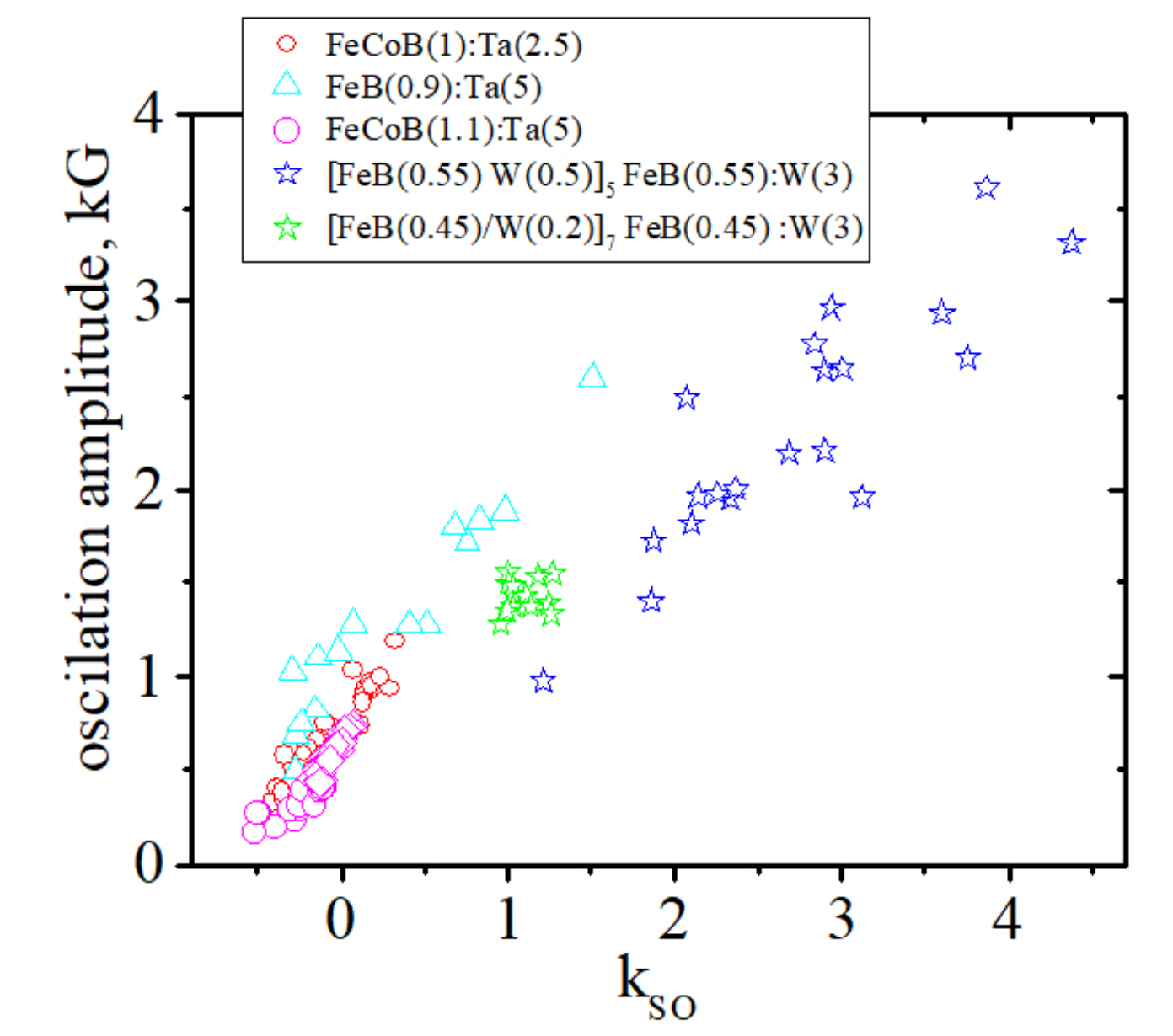}
	\end{center}
	\caption{\label{fig:fig3} 
		Oscillation amplitude of the strength of the spin-orbit interaction under an increasing external magnetic field vs. the coefficient of spin-orbit interaction. Each dot represents an individual nanomagnet measurement. 
	}
\end{figure}


Another intriguing feature observed in the measurement of SO strength is the presence of oscillations atop the linear dependence of $H_{ani}$ versus $H_z$ (See Fig.\ref{fig:fig1}), which are consistent across all measured nanomagnets. The oscillations indicate periodic variations in the strength of the spin-orbit interaction under the influence of the magnetic field $H_z$. The precise reason for this periodic modulation of SO is not fully understood. One possible explanation could be the interplay between the various influences exerted by different neighboring orbitals on the modification of the electron orbital under the $H_z$- induced  Lorentz force.

The amplitude of the oscillations decreases for thicker nanomagnets, where the bulk contribution is more prominent, clearly indicating the interface origin of the oscillations.  Figure \ref{fig:fig3} shows  the measured amplitude of the oscillations vs. $k_{so}$. The spread of data along a straight line suggests a proportional relationship between oscillation amplitude and $k_{so}$. Although the slope varies for each wafer, it consistently remains positive. The absence of a negative slope implies that the demagnetization field has no influence on the oscillations.

In conclusion, a measurement method of the strength of spin-orbit interaction has been introduced, yielding insights into this complex fundamental phenomenon. Its systematic study in FeCoB nanomagnets has unveiled distinct behaviors characterizing the spin-orbit interaction in the bulk and at interface levels, the substantial disparity in the coefficients of spin-orbit interaction between single- and multi-layer nanomagnets,  intriguing periodic oscillations in spin-orbit interaction strength under an external magnetic field, the impact of interface imperfections on both spin-orbit interaction and the demagnetization field and their competitive influence on the magnetic anisotropy.

The measurement method can provide deep experimental insights into the physical processes governing magnetic anisotropy. For instance, it reveals intriguing particularities in the relationships between the strengths of spin-orbit interaction and magnetic anisotropy affected by a gate voltage \cite{Intermag2023_SO_VCMA,MMM2022_SO_VCMA} or  electrical current \cite{MMM2020_Hani_Hoff,MMM2022_SO_SOT}  or  spin accumulation\cite{Zayets2020MishenkoSpinPol,ZayetsJMMM2023Parametric,Zayets2022IEEE_Hci}.


\bibliography{SpinOrbitBib}

\begin{thebibliography}{16}
\bibitem [{Dir()}]{Dirac}%
  \BibitemOpen
  \href@noop {} {}\bibinfo {note} {Eq.1 is a consequence of the electromagnetic
  field's invariance under relativistic transformations, also known as Lorentz
  transformations. Analogous to the electromagnetic field, the quantum field of
  electrons maintains its invariance under relativistic changes, a property
  encapsulated by the Dirac equation. Remarkably, the Dirac equation yields the
  exact same Eq. 1 for $H_{so}$. This equivalence arises from the following
  reason: The spin-orbit interaction (SO) characterizes the interaction of a
  moving electron with an electric field, a phenomenon that remains consistent
  regardless of the chosen coordinates for calculation. It is possible to carry
  out calculations in a coordinate system that moves with the electron.
  Consequently, the electromagnetic field necessitates relativistic
  transformation. Similarly, computations conducted in the steady coordinate of
  the electric field can yield the same outcome for $H_{so}$. In this scenario,
  the quantum field requires a relativistic transformation, leading to the
  deduction of SO from the Dirac Equation.}\BibitemShut {Stop}%
\end{thebibliography}

\appendix

\section{Calculation of anisotropy field when an external magnetic field is applied along the easy axis }
\label{AppendixHaniCalculation}

The equilibrium magnetization direction of a nanomagnet is the direction at which magnetic energy is the smallest. The calculation of magnetic energy is as follows:

\begin{equation}
	-E= \vec H_{total}  \cdot \vec M =(\vec H+ \vec M+
	\vec H_{demag}+\vec H_{so})  \cdot \vec M
	\label{E1st}
\end{equation}

where $\vec M$ is the magnetization, $\vec H_{total}$ is the total magnetic field inside the nanomagnet, $\vec H_{demag}$ is the demagnetization field and $\vec H_{so}$ is the magnetic field of spin- orbit interaction (SO).

The magnetic field $\vec H_{so}$ is directly related to the total magnetic field experienced by an electron. In the presence of an anisotropy, $\vec H_{so}$ can be described using the tensor $\hat{k}_{so}$ In this context, Equation \ref{HsoVsH} is modified as follows:

\begin{equation}
	H_{so}=\hat{k}_{so}(\vec H+ \vec M+
	\vec H_{demag})
	\label{HsoFields}
\end{equation}

The z-axis is established as perpendicular to the plane, while the x-axis is set within the plane.  In the case of an amorphous nanomagnet, anisotropy in spin-orbit interaction can occur only between the z- and x-axes. Consequently, the tensor $\hat{k}_{so}$ can be written as follows:

\begin{equation}
	\hat{k}_{so}=
	\begin{pmatrix}
		k_{so,x} & 0 & 0  \\
		0 & k_{so,x} & 0   \\
		0 & 0 & k_{so,z}  
	\end{pmatrix}	
	\label{ksoMatrix}
\end{equation}

The demagnetization field $H_demag$ is always directed along the z-axis and is proportional to z- component of magnetization $M_z$

\begin{equation}
	H_{demag,z}=-k_{demag} \cdot M_z
	\label{Hdemag}
\end{equation}

Rewriting Eq. \ref{E1st} in components gives

\begin{equation}
	\begin{array}{l}
		-E=M_x \cdot (H_x+M_x+H_{so,x})+ \\
		+M_z \cdot (H_z+M_z+H_{demag}+H_{so,z})
	\end{array}
	\label{E2}
\end{equation}

Substitution of  Eqs. \ref{HsoFields},\ref{ksoMatrix},\ref{Hdemag} into Eq. \ref{E2} gives
\begin{equation}
	\begin{array}{l}
		-E=M_x \cdot(1+k_{so,x}) (H_x+M_x)  +\\
		+M_z \cdot (1+k_{so,z}) [H_z+(1-k_{demag}) \cdot  M_z]
	\end{array}
	\label{E3}
\end{equation}

The effective coefficient $k_{so}$ of SO  is defined as:
\begin{equation}
	1+k_{so}=\frac{1+k_{so,z}}{1+k_{so,x}}
	\label{kso_effective}
\end{equation}

Under an applied external magnetic field the magnetization $M$ is tilted, but does not change its value. The ratio between two components of $M$ is: 

\begin{equation}
	M_z=\sqrt{M^2-M^2_x}
	\label{M}
\end{equation}

Substitution of Eqs. \ref{kso_effective}, \ref{M} into Eq. \ref{E3} gives

\begin{equation}
	\begin{array}{l}
		\frac{-E}{1+k_{so,x}}=M^2+M_x H_x+ H_z (1+k_{so}) \sqrt{M^2-M^2_x} +\\
		+(M^2-M^2_x)[(1+k_{so})(1-k_{demag})-1]
		\label{Emag2}
	\end{array}
\end{equation}

The equilibrium magnetization angle corresponds to the orientation where the magnetic energy is at its minimum. The minimum can be found from the condition:

\begin{equation}
	\frac{\partial E}{\partial M_x}=0
	\label{Emin}	
\end{equation}

When there is no perpendicular external magnetic field ($H_z=0$), the condition specified in Eq.\ref{Emin} yields:

\begin{equation}
	H_x- 2M_x[(1+k_{so})(1-k_{demag})-1]=0
	\label{Emin1}
\end{equation}
Solution of Eq. \ref{Emin1} gives the linear relation between $M_x$ and $H_x$ as:

\begin{equation}
	\frac{M_x}{M}=\frac{H_x}{H^0_{ani}}
	\label{Hani2}
\end{equation}

where the anisotropy field $H^0_{ani}$ in absence of the external field is calculated as:

\begin{equation}
	H^0_{ani}=2M[(1+k_{so})(1-k_{demag})-1]
	\label{Hani0}
\end{equation}

Substitution of Eq. \ref{Hani0} into Eq. \ref{Emag2} gives:

\begin{equation}
	\begin{array}{l}
		\frac{-E}{1+k_{so,x}}=M^2+M_x H_x+ H_z (1+k_{so}) \sqrt{M^2-M^2_x} +\\
		+(M^2-M^2_x)H^0_{ani}/(2M)
		\label{Emag3}
	\end{array}
\end{equation}

Substitution of Eq. \ref{Emag3} into condition the condition \ref{Emin} gives:

\begin{equation}
	\begin{array}{l}
		H_x \cdot (1- \frac{M_x}{M} H^0_{ani})- H_z (1+k_{so}) \frac{M_x}{\sqrt{M^2-M^2_x}}=0
		\label{Emin3}
	\end{array}
\end{equation}

The solution of Eq. \ref{Emag3}  is similar to the solution \ref{Hani0} and can be express as:
\begin{equation}
	\frac{M_x}{M}=\frac{H_x}{H_{ani}}
	\label{Hani4}
\end{equation}

where the anisotropy field $H_{ani}$ is calculated as:

\begin{equation}
	H_{ani}=H^0_{ani}+H_z \frac{1+k_{so}}{ \sqrt{1- \frac{M^2_x}{M^2}} }
	\label{Hani5}
\end{equation}

The following condition holds true for small to moderate tilting angles:
\begin{equation}
	\frac{M^2_x}{M^2} \ll 1 
\end{equation}

Thus, even in the presence of an external magnetic field, there is a linear relation between  $M_x$ and $H_x$. The anisotropy field $H_{ani}$ is calculated from Eq. \ref{Hani5} as:

\begin{equation}
	H_{ani}=H^0_{ani}+H_z+ k_{so}H_z
	\label{Hani_final}
\end{equation}

Within a nanomagnet, an internal magnetic field $H_{int}$ maintains the magnetization along its easy axis even without the external magnetic field $H_z$.  Given that the internal and external magnetic fields share the same nature, their impact on the anisotropy field should be analogous. This implies that $H_{ani}$ should solely depend on $H_{int} + H_z$. Noting that,  Eq. \ref{Hani_final} can be reformulated as:

\begin{equation}
	H_{ani}=(H_z+H_{int})+ k_{so}(H_z+H_{int})
	\label{Hani_with_Hint}
\end{equation}
where

\begin{equation}
	H^0_{ani}= H_{int}+k_{so} H_{int}
	\label{HaniVsHint}
\end{equation}

 The internal magnetic field $H_{int}$ can be calculated from Eq. \ref{HaniVsHint} as:
\begin{equation}
	H_{int}=\frac{H^0_{ani}}{1+k_{so}}
	\label{Hint}
\end{equation}

\section{ Details of fabrication and measurement }
\label{AppendixExpSetup}

The equilibrium magnetization for all examined nanomagnets was oriented perpendicular to the plane. The perpendicular magnetic anisotropy (PMA) in these nanomagnets was induced at the interface. Sufficient PMA is only observed in nanomagnets featuring a smooth interface, emphasizing the pivotal role of fabrication technology in achieving this property.

\begin{figure}[tb]
	\begin{center}
		\includegraphics[width=7 cm]{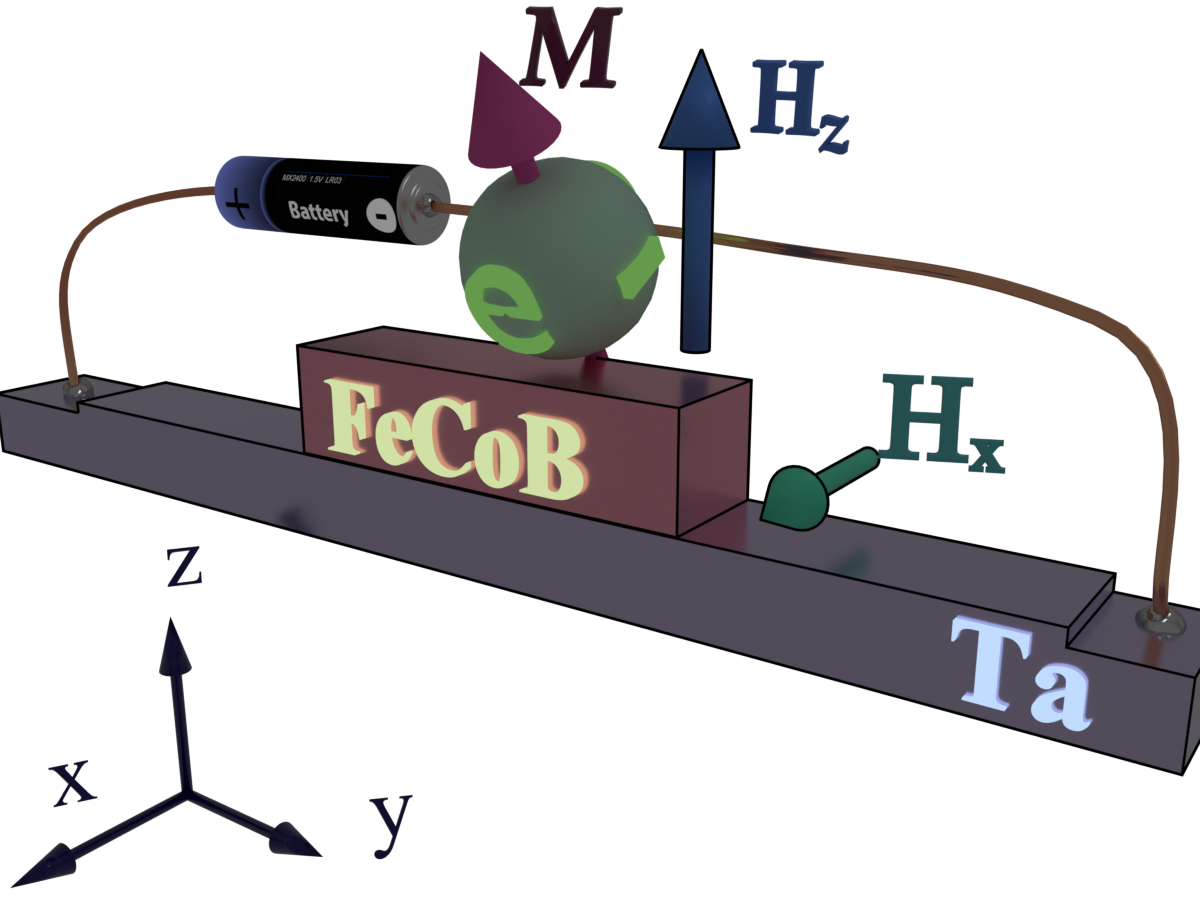}
	\end{center}
	\caption{\label{fig:figExp1} 
		Experimental setup for measuring strength of spin- orbit interaction. An external magnetic field  $H_z$  is applied along the easy axis of the FeCoB nanomagnet and used as a parameter. An additional magnetic field $H_x$ is scanned perpendicular to the easy axis.  $M$ denotes the magnetization. 
	}
\end{figure}


The sample was fabricated on a $Si$/$SiO_2$ substrate through sputtering at room temperature, followed by annealing at $T=250^0 C$. To attain a smooth surface for the nanomagnet, a buffer made of a non-magnetic metal such as tantalum or tungsten, thicker than 2 nm, was utilized. It's noteworthy that the desired perpendicular magnetic anisotropy (PMA) wasn't attained with a thinner buffer. 

The thickness of FeCoB or FeB single-layer nanomagnets ranged between 0.8 nm and 1.4 nm. Thinner nanomagnets showed either non-ferromagnetic behavior or in-plane magnetization, attributed to greater surface roughness or even film discontinuity. Thicker nanomagnets exhibited in-plane magnetization due to a significant bulk contribution, resulting in a reduced overall strength of the spin-orbit (SO) interaction (See \ref{AppendixHaniCalculation}). Multilayer nanomagnets can have a greater thickness, but increasing the number of ferromagnetic layers inevitably results in an increase of surface roughness and , as a consequence, a decrease of the SO strength and a decrease of the anisotropy field (See Fig. \ref*{fig:fig3}). 

The nanomagnet was covered by a 10 nm-thick MgO layer to induce a sufficient PMA. To shield the MgO  from air exposure, a Ta/Ru layer was used. The fabrication procedure involved multiple nanofabrication stages, employing electron-beam (EB) or optical lithography, alongside Ar milling and lift-off techniques. Each fabrication step was aligned within at least a 5 nm precision. Monitoring the etching material in real-time was carried out using a Secondary-ion-mass-spectroscopy (SIMS) detector to maintain the required etching depth accuracy.

Nanomagnets of varying sizes, ranging from 50 nm x 50 nm to 2000 nm x 2000 nm, were fabricated at different locations on a single wafer. Etching of the nanomagnet stopped at the top of the W or Ta buffer layer. Subsequently, a W or Ta nanowire with a $SiO_2$ (100 nm) isolation layer was fabricated.  A pair of Hall probes was precisely aligned with the position of the nanomagnet, and the width of the nanowire matched the width of the corresponding nanomagnet.The nanowire's etching stopped at the top of the $SiO_2$ substrate. Lastly, Cr (2 nm)/Au (200 nm) contacts were fabricated within the etched $SiO_2$, with the contact etching process ending in the middle of the Ta or W buffer layer.

\begin{figure}[tb]
	\begin{center}
		\includegraphics[width=5.5 cm]{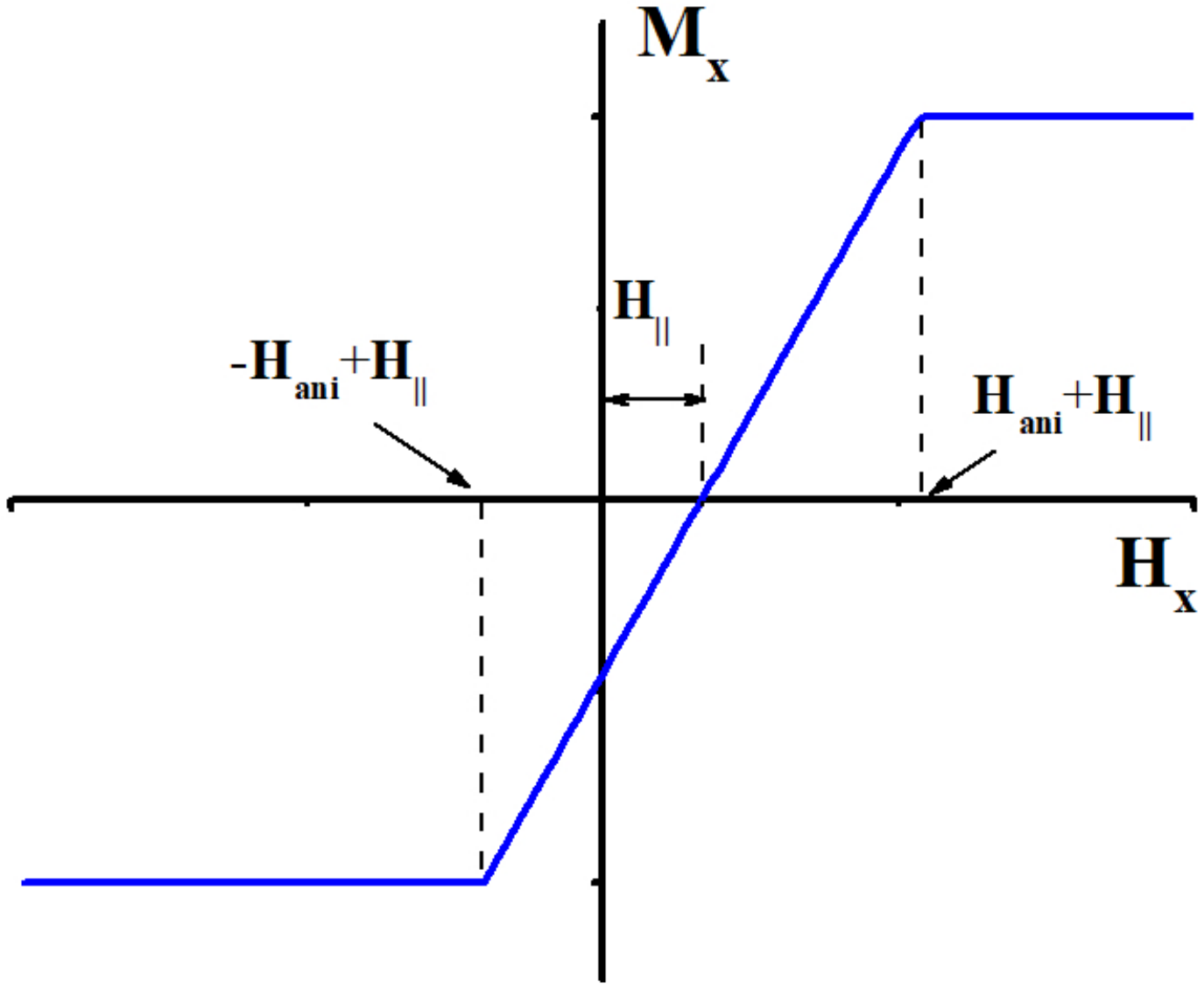}
	\end{center}
	\caption{\label{fig:figExp2} 
		Measurement principle. Schematic diagram. The in-plane component of magnetization $M_x$ vs. in-plane magnetic field $H_x$. $M_x$ is linear proportional to $H_x$ until the magnetization M aligns in-plane.		 The dependence is asymmetric versus a reversal of $H_x$ due to existence of  the intrinsic magnetic field $H_{||}$. 
	}
\end{figure}

The experiments were carried out at room temperature, well below the Curie temperature of FeCoB and FeB. A nanovoltmeter measured the Hall voltage, while a current source both supplied and gauged the electrical current. An electromagnet produced the magnetic field, offering the flexibility to align it in any direction. This allowed separate control over the in-plane and $\perp$-plane components of the applied magnetic field. To ensure accuracy, the electromagnet underwent calibration via Hall measurements conducted on non-magnetic nanowires made of Ta, W, and Ru.

The measurement procedure involved recording a measurement of the Hall angle $\alpha_{Hall}$ while sweeping an in-plane external magnetic field $H_x$ in two opposing directions. The $\perp$-plane magnetic field was used as a parameter in this measurement and kept unchanged during the  $H_x$ sweep (See Fig. \ref{fig:figExp1}).

The anisotropy field $H_{ani}$ was evaluated from the linear relation (Eq.\ref{Hani}). The Hal angle $\alpha_{Hall}$ is linearly proportional to the perpendicular- to- plane component of the magnetization $M_z$. The in- plane magnetization component $M_x$ was calculated as: 

\begin{equation}
	\frac{M_x}{M}=\sqrt{1-\left (\frac{M_z}{M} \right ) ^2}= 
	\sqrt{1-\left (\frac{\alpha_{Hall}}{\alpha_{Hall,0}} \right ) ^2}
	\label{MxVsMz}
\end{equation}

where $\alpha_{Hall,0} $ is the maximum of the Hall angle when the magnetization is perpendicular to the plane.

Figure \ref{fig:figExp2} shows the dependence of $M_x$ on $H_x$. $M_x$ is linear proportional to $H_x$ until the magnetization M is aligned in-plane. The shift in dependence \cite{Zayets2022IEEE_Hci,ZayetsJMMM2023Parametric} is a result of the presence of the in-plane magnetic field $H_{||}$. The magnetic field generated by spin accumulation and the Oersted magnetic field resulting from the current in the non-magnetic metal both contribute to this magnetic field. Disregarding the influence of $H_{||}$ could lead to a significant systematic error. To eliminate the contribution of $H_{||}$, the positive and negative segments of the measured spectrum were simultaneously fitted, allowing the determination of $H_{||}$. Subsequently, the measured spectrum was shifted based on $H_{||}$.  $M_x$ was obtained using Eq. \ref{MxVsMz}, following the evaluation of $H_{ani}$ derived from the linear fitting of $M_x$ versus $H_x$.  For each value of $H_z$, $H_x$ was systematically scanned and $H_{ani}$ was evaluated. $H_{ani}$ consistently increases as $H_z$ increases.

\end{document}